\documentclass[conference]{IEEEtran}
\usepackage{cite}
\usepackage{booktabs}
\usepackage{amsmath,amssymb,amsfonts}
\usepackage{algorithmic}
\usepackage{graphicx}
\usepackage{textcomp}
\usepackage{flushend}
\usepackage{xcolor}
\usepackage[]{mdframed}
\usepackage{subfiles} 
\def\BibTeX{{\rm B\kern-.05em{\sc i\kern-.025em b}\kern-.08em
    T\kern-.1667em\lower.7ex\hbox{E}\kern-.125emX}}
\begin{document}

\title{Code Vulnerability Detection: A Comparative Analysis of Emerging Large Language Models}
\author{\IEEEauthorblockN{Shaznin Sultana}
\IEEEauthorblockA{\textit{Department of Computer Science} \\
\textit{Boise State University}\\
Boise, Idaho, USA \\
shazninsultana@u.boisestate.edu}
\and
\IEEEauthorblockN{Sadia Afreen}
\IEEEauthorblockA{\textit{Department of Computer Science} \\
\textit{Boise State University}\\
Boise, Idaho, USA \\
sadiaafreen@u.boisestate.edu}
\and
\IEEEauthorblockN{Nasir U. Eisty}
\IEEEauthorblockA{\textit{Department of Computer Science} \\
\textit{Boise State University}\\
Boise, Idaho, USA \\
nasireisty@boisestate.edu}
}

\maketitle

\begin{abstract}
The growing trend of vulnerability issues in software development as a result of a large dependence on open-source projects has received considerable attention recently. This paper investigates the effectiveness of Large Language Models (LLMs) in identifying vulnerabilities within codebases, with a focus on the latest advancements in LLM technology. Through a comparative analysis, we assess the performance of emerging LLMs, specifically Llama, CodeLlama, Gemma, and CodeGemma, alongside established state-of-the-art models such as BERT, RoBERTa, and GPT-3. Our study aims to shed light on the capabilities of LLMs in vulnerability detection, contributing to the enhancement of software security practices across diverse open-source repositories. We observe that CodeGemma achieves the highest F1-score of 58\% and a Recall of 87\%, amongst the recent additions of large language models to detect software security vulnerabilities.
\end{abstract}


\section{Introduction}
In modern software development, reliance on open-source projects accelerates development but increases vulnerability risks. Developers can speed up their projects by taking advantage of pre-existing functionalities in these libraries, which are invaluable tools. The prevalence of open-source projects heightens the importance of monitoring security vulnerabilities within linked libraries, as these flaws can be inherited by developing products and exploited by intelligent attackers.

The number of software vulnerabilities is rapidly increasing, as shown by the vulnerability reports from Common Vulnerabilities and Exposures (CVEs) in recent years. As the number of vulnerabilities increases, there will be more possibilities for cybersecurity attacks, which can cause serious economic and social harm. Therefore, vulnerability detection is crucial to ensure the security of software systems and protect social and economic stability. 

A combination of proactive actions and technical advances is necessary to address these challenges. While helpful, traditional methods like static and dynamic analysis have their limits when it comes to finding unknown vulnerabilities in large codebases. Vulnerability identification is already a difficult task, and the complexity of today's software ecosystems, with their deep dependence networks, makes it even more so.

\begin{figure*}[t]
    \centering
    \includegraphics[width=1\linewidth]{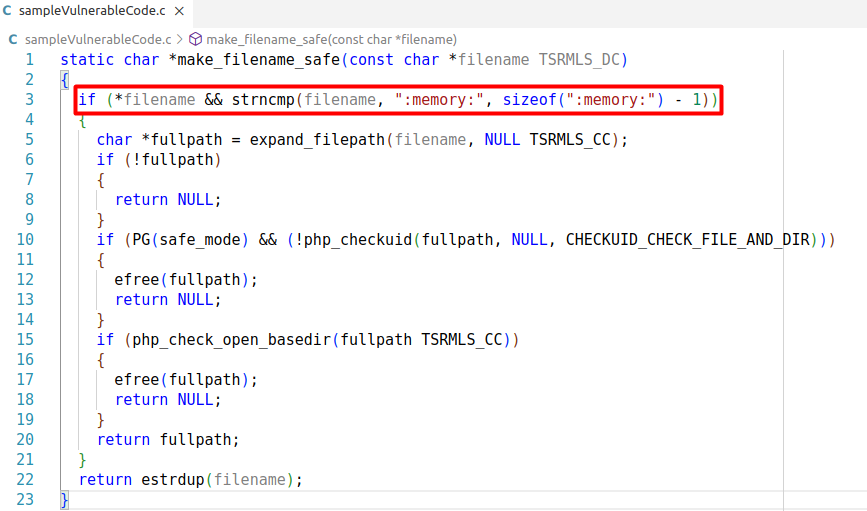}
    \caption{An example of a vulnerable code from DiverseVul dataset~\cite{chen2023diversevul}}
    \label{fig:sample-vulnerable-code}
\end{figure*}

Figure~\ref{fig:sample-vulnerable-code} is an example of a vulnerable code in C taken from DiverseVul dataset~\cite{chen2023diversevul} where it is marked as \textit{CWE-264} in project \textit{php-src} with the comment: ``Improve check for :memory: pseudo-filename in SQlite". The comment suggests an improvement to the check for the ``:memory:" pseudo-filename in SQLite. In the provided code snippet, the function make\_filename\_safe is responsible for making a filename safe for use.

The vulnerability in this code snippet (indicated with red square in the figure) lies in the comparison strncmp(filename, ``:memory:", sizeof(``:memory:") - 1). The issue here is that strncmp is used to compare the input filename with ``:memory:". However, strncmp stops comparing after a certain number of characters (specified by the third argument), which in this case is the length of ``:memory:" minus 1. This means that if the input filename is longer than ``:memory:", it won't be recognized as the ``:memory:" pseudo-filename, potentially leading to unexpected behavior or security vulnerabilities.

Recent studies~\cite{chen2023diversevul,alqarni_low_2022,chan_transformer-based_2023} have shown that LLM, and especially those in the BERT, GPT, and T5 families, are very good at detecting vulnerabilities, outperforming traditional Deep Learning models. Incredible contextual sensitivity and vulnerability indicator detection skills are displayed by these models within codebases. To extend these works, we aim to explore the capabilities of the latest additions to LLMs. We aim to further enhance the accuracy and effectiveness of code vulnerability detection in open source repositories.

The primary objective of this project is to conduct a comparative analysis of code vulnerability detection efficiency across a spectrum of LLMs. Specifically, we aim to compare the performance outcomes of less-explored LLMs, including the Llama and Gemma family of LLMs, against those of established state-of-the-art models such as BERT, RoBERTa, and GPT-3. To address these objectives and evaluate, we pose the following four research questions.
 
\begin{enumerate}
    \item \textbf{How effective are recently introduced Large Language Models in detecting code vulnerabilities?}
        \begin{itemize}
            \item Experimenting with recent LLMs (Llama 2, Gemma, CodeLlama, CodeGemma).
        \end{itemize}
        
    \item \textbf{Can natural language-based LLMs outperform code-based ones in this aspect?}
        \begin{itemize}
            \item Compare performance of Natural Language based LLMs (Llama 2, Gemma) with code based ones (CodeLlama, CodeGemma).
        \end{itemize} 

    \item \textbf{How do the findings compare with the state-of-the-art models?}
         \begin{itemize}
            \item  Compare our results with the state-of-the-art models.
        \end{itemize} 

    \item \textbf{What are the findings of these new LLMs compared to the established ones?}
        \begin{itemize}
            \item Discussion and defining criterions based on the results analysis.
        \end{itemize} 
\end{enumerate}

To this end, we fine-tune Llama 2, CodeLlama by Meta and Gemma, CodeGemma by Google with a balanced dataset curated from~\cite{chen2023diversevul}. The code base for applying these tasks are modular enough so that we can reuse the same code base for any LLMs. The reason for fine-tuning is to gain the most out of the dataset, to train the LLMs for a suitable task which in our case is source code vulnerability detection. We also create specific prompt engineered dataset from the original dataset to feed into the model. The latest LLMs are showing promising performances overall which makes us curious to verify if these can be leveraged into software vulnerability detection and show equal proficiency. Regardless, despite their impressive performance in various aspects, it is essential to critically evaluate LLMs' suitability for the specific task of software vulnerability detection.

\section{Related Works}
Over the years, much research has been conducted on different subtopics related to vulnerability detection. Hence, we have decided to collect literature on four subtopics that are most relevant to our research. This section discusses this literature, notable findings, and a few gaps that our work will exploit.

\subsection{Related Datasets}
DiverseVul~\cite{chen2023diversevul} is a dataset of vulnerable source code for C/C++, which will be used as the core dataset in this project. It is the most recent and largest dataset of vulnerable source code which accumulates 349437 vulnerable and non-vulnerable code covering 150 Common Weakness Enumerations (CWE) from a diverse set of real-world projects such as linux, vim, tcpdump, tensorflow, etc. The approach for data collection involved identifying security issue websites, parsing git commit URLs, extracting code files, and then either manually annotating popular CWEs or mapping Common Vulnerabilities and Exposures (CVE) numbers to them from the National Vulnerability Database (NVD). The paper also studied several deep learning (DL) and LLMs which used the dataset to detect vulnerability. The paper found that LLMs (RoBERTa, GPT-2, and T5 families) outperformed the state-of-the-art DL models specially with large datasets.

Chakraborty et al.~\cite{chakraborty_deep_2022} created a dataset of 1900 vulnerabilities ranging over C, C++, and Java from the Android Open Source Project, which at the time held over 1800 real-world projects. One of the concern of this project was that the compilation of the acuumulated code snippets depend on Android Manifest files of each ones which can be time-consuming to gather.

Nikitopoulos et al.~\cite{nikitopoulos_crossvul_2021} presented Cross-Vul, a dataset that contains paired vulnerable and fixed source files together which are written in more than 40 programming languages. However, due to the fact that it contained source files instead of functions, we decided not to choose it as it will take a lot of time to manually split them to functions and label them.

\subsection{Traditional Detection}
In a survey of program security vulnerabilities, Shahriar and Zulkernine~\cite{shahriar_mitigating_2012} observed that static analyzers can detect only certain types of vulnerabilities at a time and that they have limitations in coverage and programming languages as well. Another research by Ghaleb and Pattabiraman~\cite{ghaleb_how_2020} on evaluating six widely used static analyzers found that the tools cannot always detect the bugs which they were supposed to and noticed a high amount of false positive rates in all instances. An investigation led by Vassallo et al.~\cite{vassallo_how_2020} on usage of automatic static analyzers in real-life workflow found reluctance in developers. Their bugginess, need to configure in a regular basis, difficulty of tool usage are some of the common issues faced by the developers. The reluctance among developers is also confirmed by Beller et al.~\cite{beller_analyzing_2016}.

\subsection{Deep Learning based Detection}
In recent times, there has been a growing interest in the software security community to use machine learning approaches or even combining machine learning techniques with traditional approaches in finding security issues. To that end, Chakraborty et al.'s~\cite{chakraborty_deep_2022} deep learning based frameworks do significantly well, having a 34\% and 128\% increase in precision and recall, respectively. Most of the studies argue that deep learning-based security issue detection is an open problem; with a quality amount of data, it may have some application in real-world situations~ \cite{vassallo_how_2020}. Graph Neural Network (GNN) based approaches also showed promising results in a very recent study by Zhou et al.~\cite{zhou_devign_2019}. Besides GNN, LSTM, CNN, Bi-LSTMs, etc., several studies have been experimented with to detect vulnerable source codes~\cite{li_gated_2017, li_sysevr_2022, li_automated_2021,li_vuldeepecker_2018,steenhoek_empirical_2023}. 

\subsection{LLM based Detection}
In a recent systematic literature review by Hou et al.~\cite{hou_large_2023}, recent practices of using LLMs for vulnerability detection were discussed. These practices included experiments with different LLMs such as a modification of BERT specifically fine-tuned for this task, or combining sequence and graph embedding. Overall, LLMs showed promising aspects which inspires further research in this field.
Zhou et al.~\cite{zhou_large_2024} focused on exploring the effectiveness of GPT-3.5 and GPT-4 in this context, showcasing their competitive performance compared to prior methods. In-context learning (ICL) with prompts helps ChatGPT uncover software vulnerabilities without fine-tuning GPU resources. Carefully curated prompts with ICL added to the base prompt were used here to give ChatGPT task-specific insights for accurate vulnerability detection.
Mathews et al.~\cite{mathews_llbezpeky_2024} explored the utilization of LLMs for detecting vulnerabilities in Android applications by leveraging techniques such as Prompt Engineering and Retrieval-Augmented Generation. Through experimentation on the Ghera Vulnerability Dataset, the research demonstrated promising results, indicating the potential of LLMs in revolutionizing software engineering tools. The paper underscored the need for structured pipeline architectures and optimized contextual input to maximize LLMs' efficacy in vulnerability detection and remediation.

\section{Approach}
In this section, we outline our comparative study on code vulnerability detection using less explored and newest LLMs.

\begin{figure*}
\centering
\includegraphics[width=18 cm]{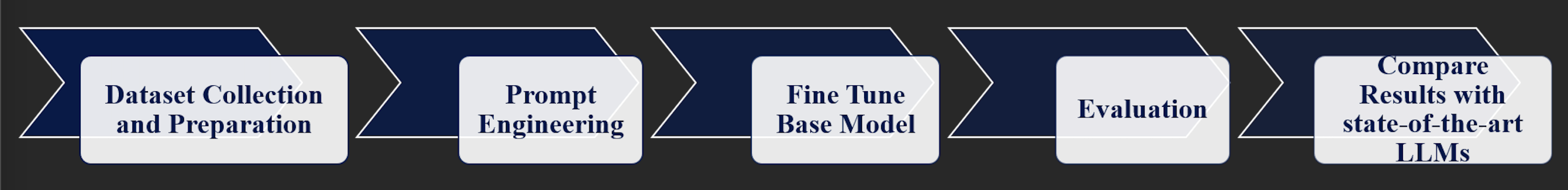}
\caption{An overall solution approach}
\label{fig}
\end{figure*}

\subsection{Dataset Collection and Preparation}
The initial step of our implementation focuses on acquiring and preparing datasets crucial for training and evaluation. We used the dataset from Chen et al.~\cite{chen2023diversevul} to perform the experiments as it is the most recent and largest dataset in this area. We applied preprocessing techniques, including tokenization, normalization, and data augmentation, to ensure dataset cleanliness and uniformity.

\begin{figure*}
\centering
\includegraphics[width=\textwidth]{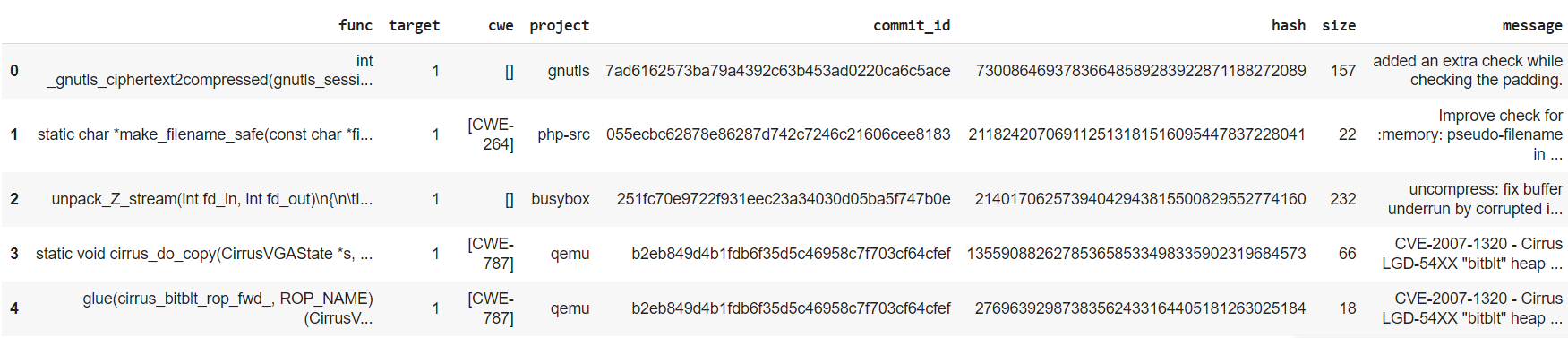}
\caption{Dataset Visualization}
\label{fig}
\end{figure*}

The dataset contains eight columns such as Function, Target, CWE, Project, Commit\_id, Hash, Size, and Message. The Function column shows 18945 code snippets that might have vulnerabilities. The target column shows if that code contains security bugs or not using `1' as vulnerable and `0' as non-vulnerable. CWE is the CWE-number representing which category of security flaw the code belongs to. Project and Commit\_id columns have a total of 797 project IDs and 7514 commits from which the codes are collected. 

\subsection{Handling Class Imbalance}
The DiverseVul dataset is a highly imbalanced dataset that has 150 CWE categories.
The imbalanced class distribution shows that there are 18945 vulnerable labeled functions and 311547 non-vulnerable. So, we used the `RandomUnderSampler' which addresses this issue by randomly removing samples from the majority class until the class distribution becomes more balanced. The parameter `sampling\_strategy' is set to 1 which means the number of samples in the minority class will be equal to the number of samples in the majority class after resampling. After under sampling, we have a balanced dataset of 37k samples out of 330k data. Then, for fine tuning, we took 1000 samples for Llama and 3000 samples for Gemma among the balanced dataset and splitted such that 80\% data to use for fine tuning and 20\% for testing.
\begin{figure}[htbp]
\centering
\includegraphics[width=\columnwidth]{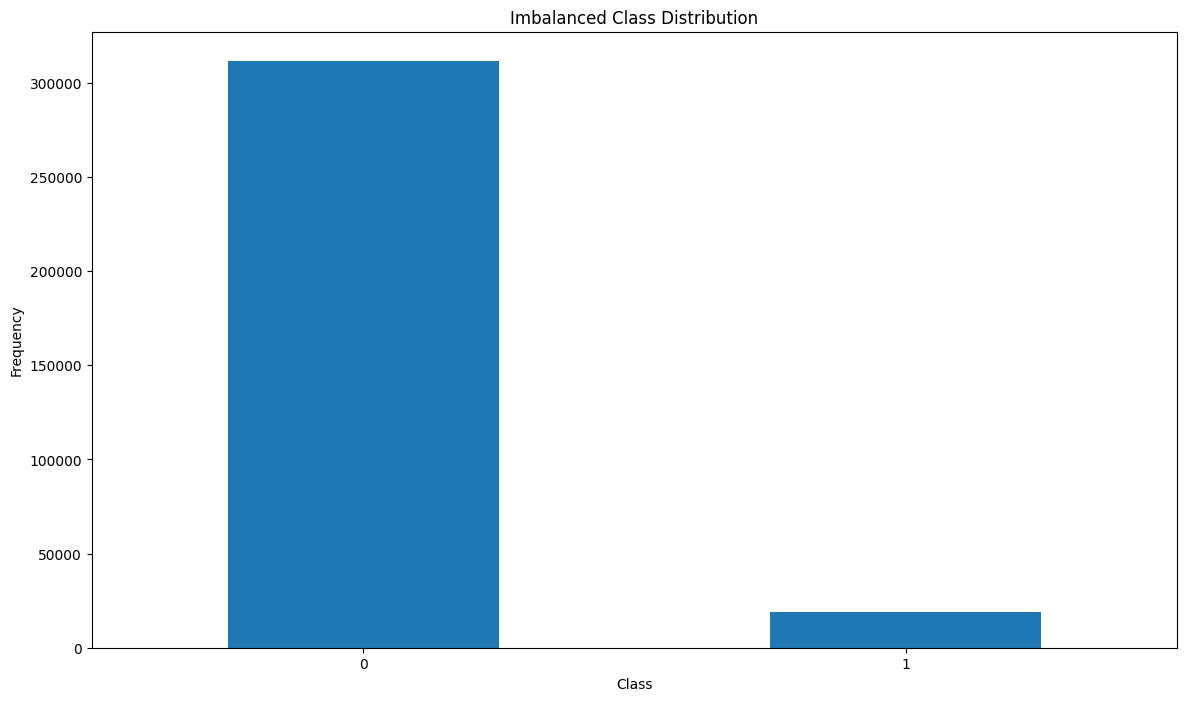}
\caption{Imbalanced Class Distribution}
\label{fig}
\end{figure}

\begin{figure}[htbp]
\centering
\includegraphics[width=\columnwidth]{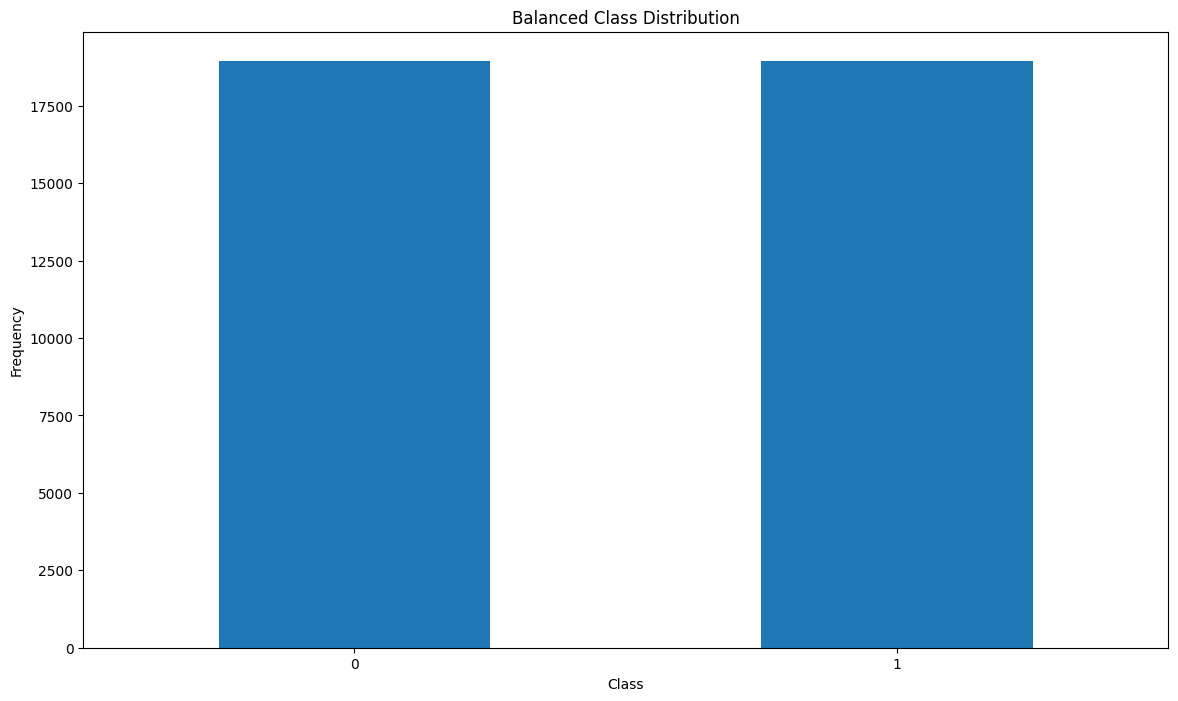}
\caption{Balanced Class Distribution}
\label{fig}
\end{figure}

\subsection{Prompt Engineering}
Prompt engineering is a method for controlling the outputs of LLMs by giving them specific instructions on what kind of data to produce. Prompt engineering relies on well-crafted prompts. The clearer, more concise, and more specific the prompt is, the more accurate the output is. The crafting process is to test, refine, and use the feedback. Llama 2 was trained with a system message that set the context and persona to assume when solving a task.

The chat version of Llama 2 requires a specific formatted prompt to generate satisfactory results.
We format our data in the specified format and engineer the prompt. The prompt engineering took several trials and errors before getting it right. Llama2 uses the One-to-Many Shot Learning Prompt Template. \textless{}s\textgreater{} token indicates the start of promting and \textless{}/s\textgreater{} is the ending. The user input starts with [INST] token and ends with [/INST].
From the models we chose, only Llama2 and CodeLlama has specific prompt format. Gemma and CodeGemma doesn't have any format to follow. So, we used the same prompt for all the models. 
Then store the formatted data for fine tuning and testing.
\begin{figure}[htbp]
\centering
\includegraphics[width=\columnwidth]{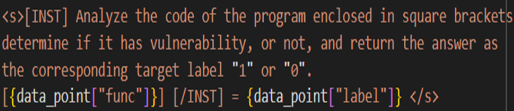}
\caption{Prompt Format}
\label{fig}
\end{figure}

\subsection{Fine Tune Base Model} 
Following dataset preparation and prompt engineering, the selected LLMs— Llama2, Gemma, CodeLlama, CodeGemma—will undergo fine-tuning and training. 

Fine-tuning procedures is designed to adapt the models' parameters to the task of code vulnerability detection while minimizing overfitting and maximizing generalization capabilities. Training is conducted using optimized hyperparameters and regularization techniques.

\begin{enumerate}
    \item Large Language Model Meta AI \textbf{Llama2}, released in July 2023 is transformer-based encoder-decoder model. LLaMA2 ranges from 7B to 65B parameters with competitive performance compared to the best existing LLMs.
    
    \textbf{CodeLlama} is code-specialized version of Llama2. Capable of generating code, and natural language about code, from both code and natural language prompts. Code Llama - Instruct, which is fine-tuned for understanding natural language instructions.
\end{enumerate}

\begin{enumerate}
    \item \textbf{Gemma}, released in February 2024 is a family of lightweight, open models built from the research and technology that Google used to create the Gemini models. These iterations of the model are trained on human language interactions and are capable of generating responses to conversational input, like a chatbot. Their compact size enables them to be deployed in environments with constrained resources.
    
    \textbf{CodeGemma} models are text-to-text and text-to-code decoder-only models and are available as a 7B pretrained variant that specializes in code completion and code generation tasks, a instruction-tuned variant for code chat and instruction following.
\end{enumerate}

\subsection{Evaluate LLMs}
Upon completion of fine-tuning, we evaluated the LLMs to assess their effectiveness in detecting code vulnerabilities. We employed a range of evaluation metrics, including accuracy, precision, recall, F1 score, and computational efficiency. We also conducted qualitative assessments to evaluate the models' ability to handle real-world code scenarios effectively.

\subsection{Compare Results with State-of-the-Art LLMs}
In the final step, we compared the performance of GPT-4, PaLM, and Llama against established state-of-the-art LLMs such as CodeBERT, CodeGPT, and GPT-2 base. This comparative analysis is to benchmark the effectiveness of the less-explored LLMs and identify areas of superiority or improvement. The comparison is based on performance metrics and computational efficiency.

\section{Implementation}
\subsection{Experimentation Setup}
As LLMs requires a large amount of memory and processing unit, we switched to Google Colab Pro to utilize the maximum resources offered by Google. We have used NVIDIA A100 GPU which is the most powerful graphics processing unit available in google colab. It also offers a System RAM of 80 GB and VRAM or GPU RAM of 40 GB with A100.
Fine-tuning each of the LLMs took 15 minutes in average, however testing took up a lot of time almost 1.5 hours for each of them. We did have to fine-tune and test 10 trials on average to get the right parameters, datapoints, prompts to receive satisfactory responses from the LLMs. We noticed a comparatively quicker time to both fine-tune and test Gemma models by Google. Figure~\ref{figllama} and \ref{figgemma} show the training loss obtained while fine-tuning the Llama 2 and Gemma models respectively. Llama required larger batch sizes than Gemma, thus the smaller number of points shown in Figure~\ref{figgemma}. It can be observed that the percentage of loss reduced as fine-tuning went on for a while. Although Gemma shows an increase in loss in comparison to Llama 2, it performs better than the latter.
\subsection{Fine-Tuning and Testing LLMs}
\begin{table*}[t]
    \centering
    \caption{Model Parameters}
    \label{tab:model_parameters}
    \begin{tabular}{lll}
        \toprule
        \textbf{Parameter} & \textbf{Gemma/CodeGemma} & \textbf{Llama 2/CodeLlama} \\
        \midrule
        bnb\_4bit\_compute\_dtype & torch.bfloat16 & torch.bfloat16 \\
        bnb\_4bit\_quant\_type & nf4 & nf4 \\
        load\_in\_4bit & True & True \\
        per\_device\_train\_batch\_size & 1 & 1 \\
        gradient\_accumulation\_steps & 4 & 1 \\
        warmup\_steps & 2 & - \\
        max\_steps & 10 & - \\
        learning\_rate & 0.0002 & 0.0002 \\
        fp16 & True & False \\
        logging\_steps & 1 & 25 \\
        optimizer & paged\_adamw\_8bit & paged\_adamw\_32bit \\
        peft\_config & lora\_config & - \\
        LoRA attention dimension & - & 64 \\
        Alpha parameter for LoRA scaling & - & 16 \\
        Dropout probability for LoRA layers & - & 0.1 \\
        use\_nested\_quant & - & FALSE \\
        bf16 & - & TRUE \\
        per\_device\_eval\_batch\_size & - & 1 \\
        gradient\_checkpointing & - & TRUE \\
        max\_grad\_norm & - & 0.3 \\
        warmup\_ratio & - & 0.03 \\
        \bottomrule
    \end{tabular}
\end{table*}
    \begin{enumerate}
    \item \textbf{Llama2 \& CodeLlama by Meta}:
    We used ‘text generation’ task model with proper prompt engineering. We selected Hugging Face’s pre trained model \textbf{“llama 2 7b chat hf”} and fine tuned. 
    As it is not possible to fully fine-tune, we use Parameter-efficient Fine Tuning (PEFT) techniques such as LoRA or QLoRA. We used QLoRa because it allows to fine-tune the model in 4-bit precision which drastically reduces the VRAM usage. 

    Some of the training arguments used for Llama2 fine tuning:

   \begin{itemize}
    \item learning\_rate = 2e-4
    \item optimizer = ``paged\_adamw\_32bit"
    \item num\_train\_epochs = 1
    \item lora\_r = 64 \# LoRA attention dimension
    \item lora\_alpha = 16 \# Alpha parameter for LoRA scaling
    \item lora\_dropout = 0.1 \% Dropout probability for LoRA layers
\end{itemize}

    We employed ‘instruction-following’ model from CodeLlama variants. We fine-tuned the Hugging Face pre-trained model \textbf{``codellama/CodeLlama-7b-Instruct-hf"}. However, it does not produce any explanation or text response like Llama2 does.

\begin{figure}[htbp]
\centering
\includegraphics[width=\columnwidth]{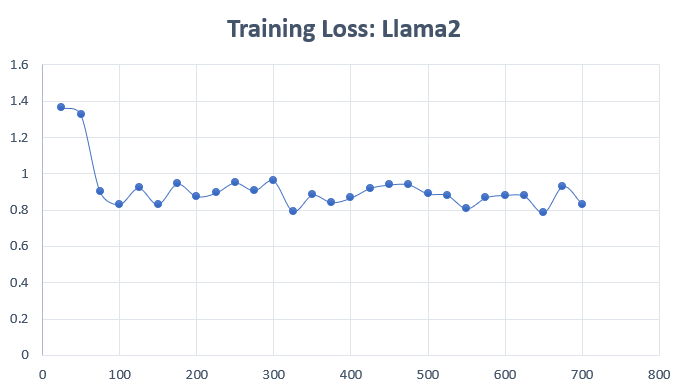}
\caption{Training Loss of Llama 2 Model}
\label{figllama}
\end{figure}

\item \textbf{Gemma \& CodeGemma by Google}:
We used the `instruct' variant for the text generation task. We selected Hugging Face’s pre-trained model \textbf{``gemma-1.1-7b-it"} and fine-tuned it.
We selected the `instruction-following' model variant for CodeGemma as well. The pre-trained model \textbf{``google/code-gemma-7b-it"} is chosen for fine-tuning. The prompt is almost the same as Llama 2, we needed to add some special tokens for processing. 
We applied Qlora technique for Gemma and CodeGemma as well.

\begin{figure}[htbp]
\centering
\includegraphics[width=\columnwidth]{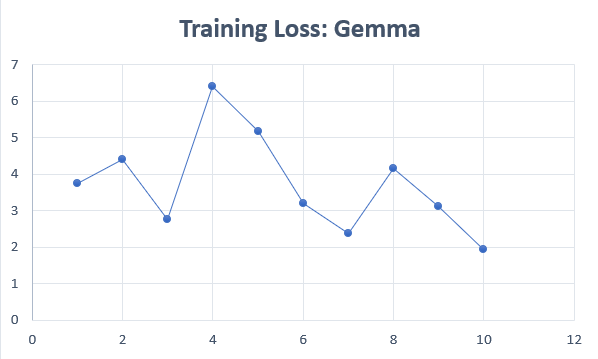}
\caption{Training Loss of Gemma Model}
\label{figgemma}
\end{figure}
\end{enumerate}

Table~\ref{tab:model_parameters} is a summary of all the parameters used to fine-tune the models, some parameters are absent for the other because each model family had their own intricacies. We evaluate the fine tuned model using a test data of 200 samples in which every sample has a prompt without label. We collected the generated text by the model for each test data item and pass it to a evaluation function. We map  `0' as nonvulnerable and `1' as vulnerable. In the real dataset, nonvulnerable data is significantly more than vulnerable data; thus, when the model produces no result, we interpret it as `0'. Finally, we use accuracy, classification report, and confusion matrix from sci kit learn library to complete the evaluation.

\section{Evaluation}
The evaluation of LLMs' performance in vulnerability detection assesses how effectively LLMs can identify security vulnerabilities in software code. For this evaluation, we measured metrics such as precision, recall, accuracy, and F1 score to determine the model's ability to correctly identify vulnerabilities while minimizing false positives and false negatives. Software vulnerabilities often exhibit complex patterns and variations, making it difficult for LLMs to generalize effectively across different vulnerabilities and software systems. 

Additionally, we will address our research questions in this section.

\subsection{RQ1 and RQ2}

To answer RQ1 and RQ2, which is comparing performance of Natural Language based LLMs with code based ones, the table `Results' is showing the metrics for each of the new LLMs performance in vulnerability detection.

\begin{figure*}
\centering
\includegraphics[width=\textwidth]{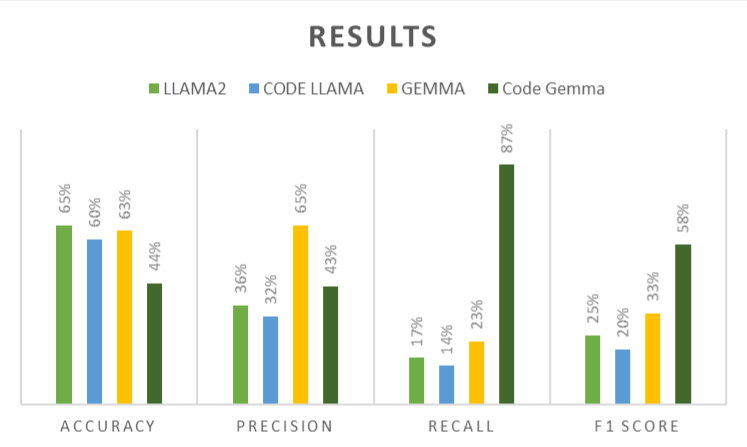}
\caption{Results}
\label{fig}
\end{figure*}

Accuracy is the measure of the overall correctness of the model's predictions. Precision indicates the proportion of true positive predictions out of all positive predictions.
Recall is the measurement of the proportion of true positive predictions out of all actual positive instances.
F1 Score is the harmonic mean of precision and recall, providing a balanced measure of a model's performance.

From the Results bar chart, we can see that Llama2 achieves highest accuracy of 65\% whereas CodeLlama is at 60\%. On the other hand, At 63\% accuracy, Gemma outperforms CodeGemma at 44\%. However, Codegemma achives highest Recall and F1 score with 87\% and 58\% respectively. For precision, Gemma is at highest with 65\% whereas Codegemma is at second with 43\%.

\begin{mdframed}
RQ1 answer: While the latest LLMs perform well overall, with the exception of accuracy, it's uncertain whether excelling in classification tasks implies equal proficiency in vulnerability detection. Strong performance in general tasks doesn't necessarily guarantee effectiveness in software vulnerability detection. It's essential to critically evaluate LLMs' suitability for the specific task of software engineering tasks.
\end{mdframed}

\begin{mdframed}
RQ2 answer: There seems to be no noticable performance differences in using natural language based LLMs and Code based LLMs except CodeGemma is doing comparatively promising performance.  
\end{mdframed}

\subsection{RQ3}

To answer RQ3, we collected performance metrics of GPT-2 Base, CodeGPT, and CodeBert from Chen et al.~\cite{chen2023diversevul}. We compare those observations with the results of our models. However, to acknowledge, the dataset for this evaluation is a merged dataset of DiversVul and two others while in our case, the dataset taken in limited constraints is only DiversVul.    

Table~\ref{tab:metrics} shows that all four of the LLMs did not produce any mentionable good results in terms of accuracy. Table~\ref{tab:metrics} is the observation when testing is on the same dataset for the models taken from the Diversvul paper. ``GPT-2 Base" and ``CodeGPT" achieve the highest accuracy of 91\%. Both of these models are traidtional ones taken from \cite{chen2023diversevul}. However, if we look into the other metrics, ``Code Gemma" achieves the highest Recall and F1 score of 87\% and 58\%, respectively, on the ``DiversVul" dataset. Also, Gemma is the highest in Precision of 65\% among the others. 

\begin{table*}[t]
  \centering
  \caption{Model Evaluation Metrics (TEST ON SAME DATASET)}
  \label{tab:metrics}
  \begin{tabular}{lccccccc}
    \toprule
    \textbf{Model} & \textbf{Dataset} & \textbf{Accuracy} & \textbf{Precision} & \textbf{Recall} & \textbf{F1 Score} \\
    \midrule
    Code Llama & Diversvul & 60\% & 32\% & 14\% & 20\% \\
    LLAMA2 & Diversvul & 65\% & 36\% & 17\% & 25\% \\
    GEMMA & Diversvul & 63\% & \textbf{65\%} & 23\% & 33\% \\
    Code Gemma & Diversvul & 44\% & 43\% & \textbf{87\%} & \textbf{58\%} \\
    GPT-2 Base & Merged Diversvul & \textbf{91\%} & 46\% & 25\% & 33\% \\
    CodeBert & Merged Diversvul & 90\% & 39\% & 36\% & 37\% \\
    CodeGPT & Merged Diversvul & 91\% & 43\% & 29\% & 35\% \\
    \bottomrule
  \end{tabular}
\end{table*}

Table~\ref{tab:metrics1} is the observation with the models from the paper in case testing is done on unseen projects. Here, the highest accuracy is still ``GPT-2 Base" with 95\%. Regardless of precision, recall, and F1 scores, the newest models are doing well, even in comparison to unseen projects.

There is another thing that should be addressed using the same merged dataset and reproducing the results could have had some differences in overall evaluation.
\begin{mdframed}
RQ3 answer: The results shows that the state of the art LLMs could outperform in Accuracy, other than that, in all the other metrics, the newest LLMs achieves better results. 
\end{mdframed}

\begin{table*}[t]
  \centering
  \caption{Model Evaluation Metrics (TEST ON UNSEEN PROJECT)}
  \label{tab:metrics1}
  \begin{tabular}{lccccccc}
    \toprule
    \textbf{Model} & \textbf{Dataset} & \textbf{Accuracy} & \textbf{Precision} & \textbf{Recall} & \textbf{F1 Score} \\
    \midrule
    Code LLAMA & Diversvul & 60\% & 32\% & 14\% & 20\% \\
    LLAMA2 & Diversvul & 65\% & 36\% & 17\% & 25\% \\
    GEMMA & Diversvul & 63\% & \textbf{65\%} & 23\% & 33\% \\
    Code Gemma & Diversvul & 44\% & 43\% & \textbf{87\%} & \textbf{58\%} \\
    GPT-2 Base & Merged Diversvul & \textbf{95\%} & 10\% & 5\% & 6\% \\
    CodeBert & Merged Diversvul & 94\% & 13\% & 10\% & 11\% \\
    CodeGPT & Merged Diversvul & 94\% & 10\% & 7\% & 8\% \\
    \bottomrule
  \end{tabular}
\end{table*}

\subsection{RQ4}

While LLMs are showing impressive capabilities, there are still some limitations and biases. It is essential to recognize those and use responsibly, taking into account the specific context and requirements of the task. LLMs are sensitive to the quality and biases present in their training data, which can lead to inaccurate or biased outputs. 

To address RQ4, we are discussing some of the interesting facts we found while evaluating the latest LLMs.

\begin{figure*}[t]
\centering
\includegraphics[width=\textwidth]{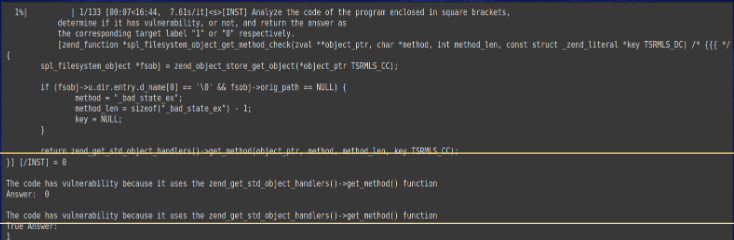}
\caption{Case 1: Llama 2 Detects vulnerability correctly with explanation}
\label{fig1}
\end{figure*}

For the first case, we can see in Figure~\ref{fig1} that Llama2 is detecting vulnerability correctly. It also gives and explanation behind this code snippet not having security issues. Besides, according to the prompt given, it gives answer in `0' or `1'. In this case, it shows '0', which means non-vulnerable. This is actually the ideal scenario.

\begin{figure*}[t]
\centering
\includegraphics[width=\textwidth]{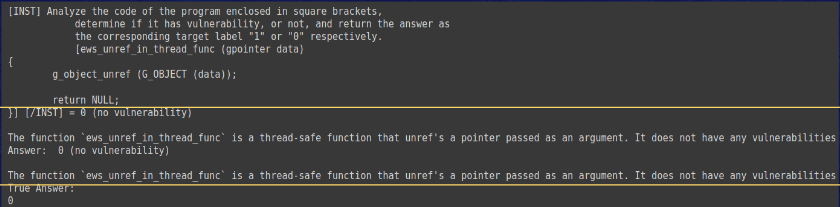}
\caption{Case 2: Llama 2 Detects vulnerability with explanation but gives response ``0"}
\label{fig2}
\end{figure*}

However, in some cases, it does not follow the ideal way. Figure~\ref{fig2} shows this issue as the second case. So, here in the figure, we can see that Llama2 detects vulnerability correctly and gives appropriate reasoning behind the answer. But the answer given according to the prompt is the opposite of the explanation, which is `0'. The answer should be `1' meaning vulnerable code detected. This clearly contradicts the LLM's capabilities and, at the same time, affects the performance metrics.

In the third case, there is another issue with another LLM, which is Gemma. This case in Figure~\ref{fig2} shows that Gemma is detecting vulnerability and gives explanations but doesn't follow the prompt instructions. So, it responds with only an explanation and no `0' or `1' value. This kind of problem are not in every case but significantly questions the overall performance. 

\begin{figure}[htbp]
\centering
\includegraphics[width=\columnwidth]{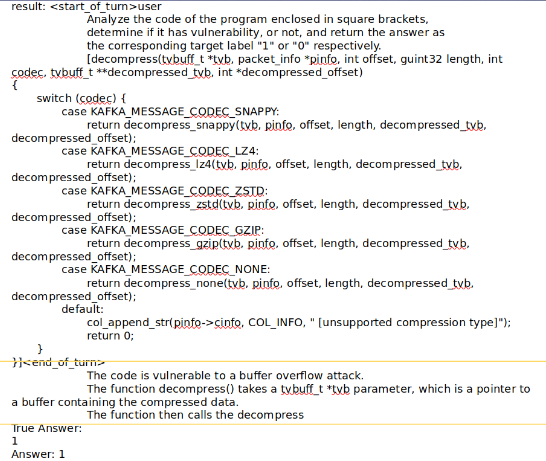}
\caption{Case 3: Gemma Detects vulnerability with explanation not ‘0’ or ‘1’}
\label{fig3}
\end{figure}

\begin{mdframed}
RQ4 answer: The results show that the LLMs are unpredictable. The prompt can still not be as useful as expected in every case. To summarize, these new LLMs are showing promising results, but there are still issues to be fixed when using them in specific fields or tasks. 
\end{mdframed}

\section{Challenges}
This research endeavor has faced several challenges.
LLMs have different frameworks and methods, so different approaches needed to bring together to make a single framework. There is not a lot of research on the exact combination of datasets and models proposed in our work, even though there is a lot of interest in LLMs for detecting code vulnerabilities. Moreover, the complexity of LLMs, characterized by their vast size and multitude of parameters, poses technical challenges in implementation and fine-tuning for our specific dataset.

We partially resolved that, and we had to change the dataset and prompts for different kinds of LLMs to formulate a code module that will work for them.
To find the perfect combinations of parameters and tweaking to fine-tune, the Llama models were fine-tuned on just 700 data, whereas Gemma models were fine-tuned on roughly 2000 data. To decode the answer generated by LLMs to a well-formatted one, CodeGemma failed to produce a well-fitted `0' or `1' answer, and we had to manipulate the strings of output to resolve.
Moreover, as LLMs need high computations, the project was carried out using Colab Pro. At first we had plan to utilize the University's high computing resources: Borah computing cluster, however, after several challenges, we had to skip that and move to Colab Pro. Also, working with LLMs require immense amount of data and to process them need expensive and high performance experiment setup. LLMs' computational demands may limit their scalability in processing large codebases or real-time applications.
Overall, Addressing the challenges of working with LLMs require a combination of technical expertise, ethical considerations, and careful experimentation to ensure the effective and responsible use of LLMs in various applications.

\section{Conclusion}

The growing dependence on open source software has expedited development processes but also heightened the risks associated with vulnerabilities, necessitating vigilant monitoring of various risks. Recent studies have emphasized the effectiveness of Large Language Models (LLMs), particularly those belonging to the BERT and GPT families, in detecting vulnerabilities. To extend these works, we explored the capabilities of the latest additions to LLMs in software engineering, potentially improved in accuracy, efficiency, and generalization.

As part of the conclusion, we want to address some of our realizations and future works. We learned that LLMs can not be evaluated with just some known strong metric performances. While an LLM might excel in a specific task, assuming similar performance in related tasks can lead to unexpected results. Each task presents unique challenges and nuances that may not be adequately addressed by the model's capabilities. Therefore, it is important to evaluate an LLM's performance in each task and field independently rather than assuming uniform proficiency across similar tasks. For future work, we could use the larger versions of these LLMs with merged and larger dataset. Also, we acknowledge that the results we have taken from the paper~\cite{chen2023diversevul} should have been reproduced with only the Diversvul dataset. 
Overall, balancing LLMs' performance with computational resources and deployment considerations in software engineering environments is essential to ensure efficiency and scalability.

\bibliographystyle{IEEEtran}
\bibliography{refs}
\end{document}